\documentclass[preprint2]{aastex}
\usepackage{graphicx}

\shortauthors{Hopkins et al.}
\shorttitle{Star Formation in Galaxies between $0.7<z<1.8$}

\begin{document}

\title{Star Formation in Galaxies between $0.7<z<1.8$}

\author{A. M. Hopkins\altaffilmark{1}, A. J. Connolly\altaffilmark{2},} 

\affil{Department of Physics and Astronomy, University of Pittsburgh, 
  3941 O'Hara Street, Pittsburgh, PA 15260, USA}

\and

\author{A. S. Szalay}

\affil{Department of Physics and Astronomy, The Johns Hopkins University, 
  3400 N. Charles Street, Baltimore, MD 21218, USA}

\altaffiltext{1}{Email: ahopkins@phyast.pitt.edu}
\altaffiltext{2}{Email: ajc@phyast.pitt.edu}

\begin{abstract}
We present an analysis of the star formation rate in galaxies between
$0.7<z<1.8$ using Near Infrared Camera and Multi-Object Spectrograph
(NICMOS) grism spectral observations. We detect 163 galaxies in
an area of $\sim 4.4$ square arcminutes, 37 of which show possible
H$\alpha$ emission. We extend the observed H$\alpha$ luminosity function (LF)
in this redshift range to luminosities a factor of two fainter than
earlier work, and are consistent in the region of overlap.
Using the H$\alpha$ LF, we estimate a star formation rate (SFR) density in
this redshift range of 0.166\,M$_{\odot}$yr$^{-1}$Mpc$^{-3}$
($H_0=75\,$km\,s$^{-1}$Mpc$^{-1}$), consistent with other estimates
based on emission lines, and supporting the order of magnitude increase in
SFR density between $z=0$ and $z=1$. Our measurement of SFR density is a
factor of $\approx 2-3$ greater than that estimated from UV data, comparable
to the factor observed locally, implying little evolution in the relative
extinctions between UV and H$\alpha$ out to $z\approx1.3$.
 
\end{abstract}


\section{Introduction}
\label{int}

Recent calculations based on measurements of the ultra-violet (UV) continuum
\citep[280\,nm,][]{Lil:95,Con:97}, [O{\sc ii}] line emission \citep{Hog:98},
and H$\alpha$ line emission \citep{Gla:99,Yan:99}, 
have shown that the comoving star formation rate (SFR) density
in galaxies increases rapidly as a function of redshift, rising by an
order of magnitude from $z=0$ to $z=1$. Measurements of the UV
luminosity of high redshift ($2<z<4$) galaxies by \citet{Ste:99} and
\citet{Mad:96}, even after correction for extinction due to dust obscuration
\citep{Ste:99}, suggest a flattening or a decrease in this trend.
Consequently, the redshift range $1<z<2$ appears to host a change
in the slope of the global SFR density, and has become the focus of
several recent studies \citep{Yan:99,Con:97}.

We have undertaken a study of the global SFR within this redshift range
using the H$\alpha$ line emission as a probe of star formation, detected
in observations with the Near Infrared Camera and Multi-Object Spectrograph
(NICMOS) on the Hubble Space Telescope (HST).
Our target regions were chosen to lie within the Groth-Westphal strip
\citep{Gro:94}. This allows us to exploit existing deep multicolour photometry
and ongoing complementary ground-based spectroscopy in the analysis
of the detected galaxies. The observations and data processing are described
in the next section, with the resulting catalogue and measured SFR density
presented in Section~\ref{res}.

Throughout this paper we assume $H_0=75\,$km\,s$^{-1}$Mpc$^{-1}$
and $q_0=0.5$.

\section{Observations and Data Processing}
\label{obs}

Field selection was based on photometric redshift estimates for sources in
the Groth-Westphal strip. To ensure detection of the
$1 \lesssim z \lesssim 2$ emission line population, the target
fields were chosen to maximise the number of galaxies
with photometric redshifts $z_{p}\gtrsim1$.

Six such fields lying within the Groth-Westphal strip
were observed with camera 3 on NICMOS using the F160W broadband filter
and the G141 grism. The grism covers the wavelength range 1.1 to
1.9\,$\mu$m with a mean dispersion of $8.1\times10^{-3}\,\mu$m pixel$^{-1}$.
More details of the grism properties are given in \citet{McC:99}.

The observations for each of the six fields consisted of 10 separate
exposures with both the grism and broadband filter. The total exposure
time with the F160W filter was $\sim 1000$\,s, and with the G141 grism was
$\sim 6200$\,s for each field. The size of the direct images is
256$\times$256 pixels with $0\farcs2$\,pixel$^{-1}$, so the six fields cover
a total area of $\sim 4.37$ square arcminutes. For all but one of
the fields the instrument was rotated by a few degrees after the first
6 pairs of exposures. This was done in an attempt to regain information
on sources whose spectra might be heavily blended at the initial position
angle.

The data processing performed closely follows the steps outlined by
\citet{McC:99}. The starting point was the calibrated output
from the STScI data processing pipeline, which
consists of dark current removal, linearisation, bad pixel masking
and cosmic-ray rejection. Subsequent processing began with subtraction of
a median sky frame. The median sky was constructed by combining
the calibrated exposures for all the observations, producing a high
signal-to-noise (S/N) sky frame. This median was scaled before subtraction
from the grism images to minimise the rms noise in each frame. The individual
sky-subtracted frames for each field were then combined using a median filter
after being shifted into registration using offsets derived from the
associated direct images. The observations at different position angles
for each field were processed independently, resulting in two direct
and grism images (at different position angles) for five of the six fields.

The NICMOS-Look software \citep{Fre:98} was used for the extraction of
one-dimensional spectra from the grism images. The location and sizes for the
objects in each frame were established using SExtractor \citep{BA:96}.
The resulting source lists were input to NICMOS-Look for spectral extraction.
This procedure identifies extraction regions in the grism images for each
object based on previously calibrated transformations from the object
positions in the associated direct images. The extracted spectra were
wavelength and flux calibrated based on existing calibration data incorporated
into NICMOS-Look. For some objects where the spectrum was mildly contaminated
by that of a nearby object, the ``deblend" option of NICMOS-Look was
successfully applied. For situations with strong blending, however, the
spectra were not able to be reliably extracted
\citep[more details of the ``deblend" algorithm are given in ][]{Fre:99}.
In the latter cases unblended spectra were often available in the
grism image taken at the other position angle. In some few cases,
the spectra of objects in more crowded regions were heavily blended at both
position angles, and no reliable spectra for these were obtained.
For these reasons, spectra for corresponding objects in the pairs of
rotated images were not combined. Instead the spectra were analysed
separately. In most cases the parameters for detected emission lines
were measured using the spectrum extracted from the image at the
orientation with the higher S/N.

While the spectral range of the grism is 1.1 to $1.9\,\mu$m,
we found that in many spectra, at wavelengths longer than $\sim 1.85\,\mu$m,
spurious features or artifacts prevented confident emission line detections.
As a result we have restricted our analysis to $z \leq 1.8$, this upper limit
corresponding to H$\alpha$ emission at $1.84\,\mu$m.
The resulting calibrated spectra show noise levels between $1-2\,\mu$Jy
(rms) over the wavelength range $1.1-1.9\,\mu$m. For a $2\sigma$
emission feature of average equivalent width ($\sim 230\,$\AA,
estimated from observed emission lines), this translates to
$S_{H\alpha}=1.66\times10^{-19}\,$W\,m$^{-2}$,
$L_{H\alpha}=2.00\times10^{34}\,$W, and
SFR$_{H\alpha}=1.6\,$M$_{\odot}$yr$^{-1}$ at $z=0.7$,
or $S_{H\alpha}=6.13\times10^{-20}\,$W\,m$^{-2}$,
$L_{H\alpha}=5.96\times10^{34}\,$W, and
SFR$_{H\alpha}=4.7\,$M$_{\odot}$yr$^{-1}$ at $z=1.8$.

\section{Results}
\label{res}

\begin{deluxetable}{ccccccc}
\tablewidth{0pt}
\tablecaption{Galaxies with possible H$\alpha$ emission. 
 \label{gals}}
\tablehead{
\colhead{Name} & Note\tablenotemark{c} & \colhead{$S_{H\alpha}$} & Eq. Width & \colhead{$z$} & \colhead{$L_{H\alpha}$} & SFR \\
\colhead{(J2000 coordinates)} & & \colhead{($\times 10^{-19}$\,W\,m$^{-2}$)} &\colhead{(\AA)} & \colhead{} &\colhead{($\times 10^{34}$\,W)} & \colhead{(M$_{\odot}$yr$^{-1}$)}
}
\startdata
NIC~J141725.37+522721.0 & 1 & 0.65 & 179 & 1.31 & 2.20 & 1.7 \\
NIC~J141725.66+522652.3 & 3 & 2.37 & 244 & 1.70 & 14.4 & 11.3 \\
NIC~J141725.66+522724.7 & 1 & 0.83 &  66 & 1.07 & 1.79 & 1.4 \\
NIC~J141725.71+522652.9 & 3 & 3.49 & 257 & 1.71 & 21.4 & 16.8 \\
NIC~J141726.75+522449.0\tablenotemark{a} & & 2.14 &  68 & 0.83 & 2.61 & 2.1 \\
NIC~J141727.28+522637.4\tablenotemark{a} & & 3.43 & 223 & 1.00 & 6.36 & 5.0 \\
NIC~J141727.34+522647.0\tablenotemark{a} & & 3.51 & 381 & 1.55 & 17.2 & 13.5 \\
NIC~J141727.42+522644.5\tablenotemark{a} & & 2.50 & 198 & 0.91 & 3.79 & 3.0 \\
NIC~J141727.46+522445.0 & 2 & 1.27 & 253 & 0.99 & 2.32 & 1.8 \\
NIC~J141729.60+522446.4\tablenotemark{a} & & 2.46 & 367 & 0.97 & 4.28 & 3.4 \\
NIC~J141740.30+522645.0\tablenotemark{a} & & 1.17 & 162 & 1.04 & 2.37 & 1.9 \\
NIC~J141740.65+522648.8 & 3 & 1.69 & 125 & 0.92 & 2.60 & 2.0 \\
NIC~J141740.71+522659.8\tablenotemark{a} & & 1.70 &  69 & 0.75 & 1.68 & 1.3 \\
NIC~J141741.93+522645.2\tablenotemark{a} & & 2.96 & 168 & 0.75 & 2.97 & 2.3 \\
NIC~J141750.32+523054.3 & 3 & 3.54 & 333 & 1.12 & 8.42 & 6.6 \\
NIC~J141750.88+523043.7 & 2 & 1.66 & 271 & 0.95 & 2.75 & 2.2 \\
NIC~J141751.06+523040.0 & 3 & 2.55 & 201 & 1.55 & 12.6 & 9.9 \\
NIC~J141751.10+523023.6 & 2 & 1.43 & 152 & 1.22 & 4.15 & 3.3 \\
NIC~J141751.15+523049.3 & 1 & 2.28 & 344 & 1.00 & 4.24 & 3.3 \\
NIC~J141751.99+523052.3\tablenotemark{a} & & 1.10 &  61 & 1.00 & 2.06 & 1.6 \\
NIC~J141752.29+523008.3\tablenotemark{a} & & 1.65 & 280 & 1.10 & 3.79 & 3.0 \\
NIC~J141753.47+522922.2\tablenotemark{a} & & 4.13 & 451 & 1.00 & 7.76 & 6.1 \\
NIC~J141753.77+523033.3\tablenotemark{a} & & 3.61 & 298 & 1.00 & 6.80 & 5.4 \\
NIC~J141754.00+522923.9\tablenotemark{a} & & 1.27 & 164 & 0.99 & 2.33 & 1.8 \\
NIC~J141754.17+522907.3\tablenotemark{a} & & 0.97 & 161 & 0.99 & 1.77 & 1.4 \\
NIC~J141754.23+523042.6\tablenotemark{a} & & 2.58 & 199 & 0.81 & 2.98 & 2.3 \\
NIC~J141754.29+523023.0 & 2 & 1.65 & 301 & 1.43 & 6.79 & 5.3 \\
NIC~J141754.30+522916.3 & 2 & 1.06 & 217 & 1.53 & 5.05 & 4.0 \\
NIC~J141754.35+523036.5\tablenotemark{b} & 2 & 0.89 & 209 & 1.24 & 2.66 & 2.1 \\
NIC~J141754.54+523015.1\tablenotemark{a} & & 2.52 & 401 & 1.63 & 13.9 & 10.9 \\
NIC~J141754.55+522904.2 & 2 & 0.58 & 180 & 0.93 & 0.935 & 0.74 \\
NIC~J141754.57+522930.6\tablenotemark{a} & & 0.82 & 249 & 1.00 & 1.52 & 1.2 \\
NIC~J141755.24+522928.2 & 2 & 1.53 & 253 & 1.34 & 5.47 & 4.3 \\
NIC~J141757.52+522922.5 & 1 & 2.89 & 278 & 0.92 & 4.49 & 3.5 \\
NIC~J141757.94+522926.8 & 2 & 1.11 & 249 & 1.25 & 3.42 & 2.7 \\
NIC~J141759.68+523023.4 & 1 & 0.42 & 183 & 1.55 & 2.06 & 1.6 \\
NIC~J141800.61+523011.4 & 2 & 0.93 & 188 & 1.27 & 2.93 & 2.3 \\
\enddata
\tablenotetext{a}{These objects have ground based optical spectra confirming
the redshift (the DEEP consortium, private communication).}
\tablenotetext{b}{This object has a ground based photometric redshift
consistent with this emission line being H$\alpha$.}
\tablenotetext{c}{Confidence level of possible H$\alpha$ emission.
1: low confidence; 2: mid-confidence; 3: high confidence. For
galaxies with ground-based spectroscopic redshifts this estimator is omitted.}
\end{deluxetable}

\begin{deluxetable}{cccc}
\tablewidth{0pt}
\tablecaption{Schechter function parameters for H$\alpha$ luminosity
functions.
 \label{schechpar}}
\tablehead{
\colhead{} & \colhead{$\phi^{*}$} & \colhead{$\log(L^{*})$} & \colhead{$\alpha$} \\
\colhead{} & \colhead{(Mpc$^{-3}$)} & \colhead{(W)} & \colhead{}
}
\startdata
Fig.~\ref{hlf}(a) & $10^{-3.48\pm0.33}$ & $36.02\pm0.23$ & $-1.86\pm0.14$ \\
Fig.~\ref{hlf}(b) & $10^{-2.54\pm0.20}$ & $35.55\pm0.11$ & $-1.60\pm0.12$ \\
\enddata
\end{deluxetable}

A total of 163 galaxies were detected in the 6 fields observed.
Our direct imaging observations are complete to m$_{AB}$(H)\,$=23.5$.
although galaxies as faint as m$_{AB}$(H)\,$\approx25$ are detected.
Two examples of the NICMOS direct and grism images showing
emission line galaxies are presented in Figures~\ref{imgs1} and \ref{imgs2}.
Of the catalogued sources, 37 show spectral features which are possibly
H$\alpha$ emission. The {\sc iraf} task {\sc splot}
was used to determine the emission line parameters. The resulting line
fluxes, equivalent widths, and redshifts are shown in Table~\ref{gals},
along with a confidence parameter, from 1 (low confidence) to
3 (high confidence), assigned subjectively as an indicator of our
confidence in the reality of the emission.

The H$\alpha$ flux, $S_{H\alpha}$, is calculated assuming that H$\alpha$
contributes 0.71 of the blended H$\alpha$/[N{\sc ii}] line \citep{McC:99}.
Also shown in Table~\ref{gals} are the derived H$\alpha$ luminosities
and estimated star formation rates (assuming that there is no AGN
contribution to the emission). The SFRs are calculated using the calibration
SFR$_{H\alpha}\,($M$_{\odot}$yr$^{-1}) = L_{H\alpha}\,$(W)$/1.27\times10^{34}$
taken from \citet{Ken:98}.

Of the 37 candidate H$\alpha$ emission line galaxies, the grism redshifts
for 17 objects (46\%) are supported by independent ground-based optical
spectroscopy from the DEEP consortium, confirming that the emission
line detected in the grism observations is indeed H$\alpha$. For a
subset of these 17 objects, further confirmation is provided by CFRS
spectroscopy \citep{Lil:95,Bri:98} for 5 objects, and for 2 more by
\citet{Koo:96}.

The discrepancy between the apparent surface densities of H$\alpha$ candidates
detected here (37 sources in $\approx 4.4$ square arcminutes) and in
the survey by \citet{Yan:99} (33 sources in $\approx 65$ square arcminutes)
is notable but misleading, and is briefly discussed here. The primary
cause of the discrepancy comes from the widely varying sensitivities
between the different fields observed in the latter survey
\citep{McC:99,Yan:99}, which comprises a large number of less sensitive
fields, and fewer deeper fields. The effect of these heterogeneous detection
levels is to produce a sample biased even more towards brighter
objects than had the observations been homogeneously flux density limited.
A prediction for the number of sources expected by integrating the
luminosity function (LF) derived by \citet{Yan:99} emphasises this result.
The number of H$\alpha$ sources expected in the current survey predicted
from this LF is slightly greater than the number we detect, after correcting
our areal coverage to account for the fact that the grism is only
sensitive to the full range of wavelengths over a portion of its area
\citep{Yan:99}. This result is not unexpected, as there will be some
incompleteness in our detections at the low luminosity end
(since we are not sensitive to the faintest luminosity
sources over the full redshift range sampled). We emphasise that
our detection rate for H$\alpha$ candidates is consistent with
the LF estimated by \citet{Yan:99}.

\subsection{Luminosity functions and density of star formation}
\label{lfs}

The H$\alpha$ luminosity function (LF) has been calculated using the
$1/V_{\rm max}$ method, with $V_{\rm max}$ as given
by \citet[ their Equation~1]{Yan:99}, having a minimum as well as a
maximum redshift being defined by the NICMOS G141 grism spectral window.
This has been done for two cases, using
(1) only galaxies with spectroscopic redshifts or confident, high S/N
detections; and (2) all galaxies with possible H$\alpha$ emission;
effectively providing lower and upper limits, respectively,
to the observed $L_{H\alpha}$ density. The resulting LFs are shown in
Figure~\ref{hlf}. The error bars are the square roots of the variances
(the sum of the squares of the inverse volumes) for each point on the
LF. Schechter function parameters (given in Table~\ref{schechpar}) were
estimated from a minimum $\chi^2$ fit, using our H$\alpha$ LF data for
each of the above two cases, combined in each case with the data of
\citet{Yan:99}. The uncertainties in our estimated parameters were
obtained using a Monte-Carlo method, based on fitting Schechter functions
to a large number of simulated H$\alpha$ LFs. The simulated LFs were
constructed with errors having a Gaussian distribution based on
the uncertainties of our measured LF points. The quoted uncertainties in
Table~\ref{schechpar} are the $1\,\sigma$ values of Gaussian fits to the
resulting distributions of each of the Schechter function parameters
independently \citep[c.f.][]{Gal:95}. It is well known, however, that
these parameters are not independent \citep[e.g.][]{Gal:96}, and this
can be seen in Figure~\ref{onesig}, which shows the $1\,\sigma$
``error-areas" for pairs of Schechter parameters from the Monte-Carlo
analysis in the second of our two cases.

In comparison to our $z\approx1.3$ results, the Schechter function parameters
for the H$\alpha$ LF of the local Universe measured by \citet{Gal:95}
(converted to $H_0=75\,$km\,s$^{-1}$Mpc$^{-1}$) are
$\phi^*=10^{-3.07}$\,Mpc$^{-3}$, $L^*=10^{34.80}\,$W, and $\alpha=-1.3$.
While the evolution of the H$\alpha$ luminosity function from
these values is clearly demonstrated in Figure~\ref{hlf}, the Schechter
parameters from our fitting appear rather extreme. In particular, we
derive low normalisations, bright $L^*$ values and steep faint end slopes,
which are strongly at variance with earlier estimates for emission line
LFs around $z\approx1$. This is due to the fact that we are examining
two cases comprising a lower and upper limit to the LF. The resulting
Schechter functions should not be taken to reflect the true shape of
the LF, although their integrals, being more robust \citep{Gal:96},
provide valid limits on our estimate of the total H$\alpha$ luminosity
density.

The total SFR density has been calculated from the integral of the
H$\alpha$ luminosity functions, and is shown
compared with emission line and UV-continuum derived estimates from
the literature in Figures~\ref{sfd1} and \ref{sfd2}. The values
in Figure~\ref{sfd1} show the UV based points uncorrected
for extinction. Figure~\ref{sfd2} applies the extinction corrections used in
\citet{Ste:99}, derived from the \citet{Cal:97a,Cal:97b} reddening law, to the
UV based estimates. This raises these points to values consistent with
the emission line derived values, and emphasizes the importance of extinction
corrections. However, \citet{Ste:99} caution against reading too
much into the high-redshift end of the latter figure (c.f. their Figure~9)
due to both the uncertainty in the faint end of the UV luminosity function at
high redshifts, and the uncertainty in the dust extinction at all redshifts.

\subsection{Individual galaxies}

Some comments on selected galaxies follow. The naming convention for
sources follows IAU recommendations, and uses J2000 coordinates.

\begin{figure*}
\centerline{\hfill
\rotatebox{0}{\includegraphics[width=7cm]{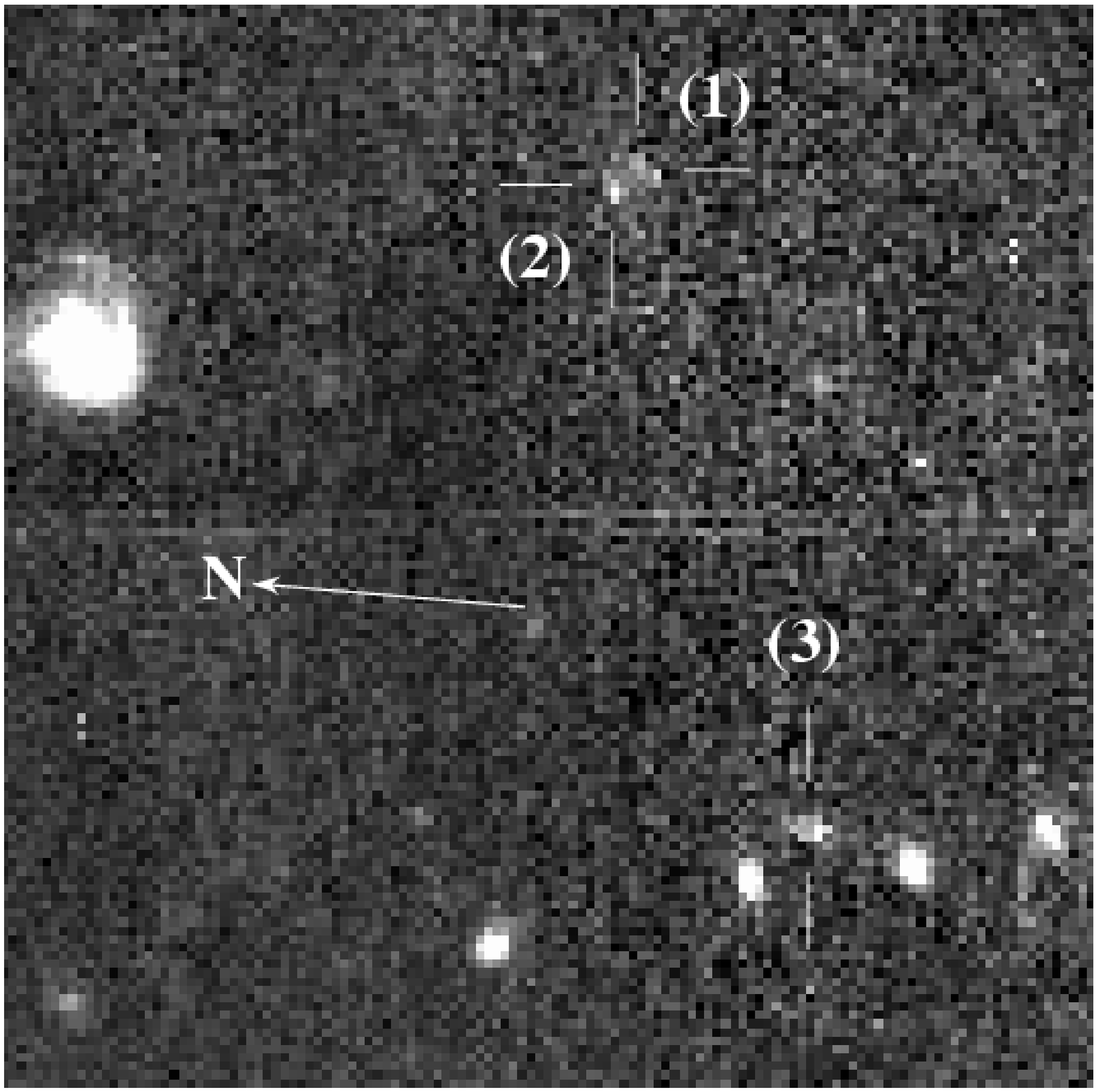}}
\hfill
\rotatebox{0}{\includegraphics[width=7cm]{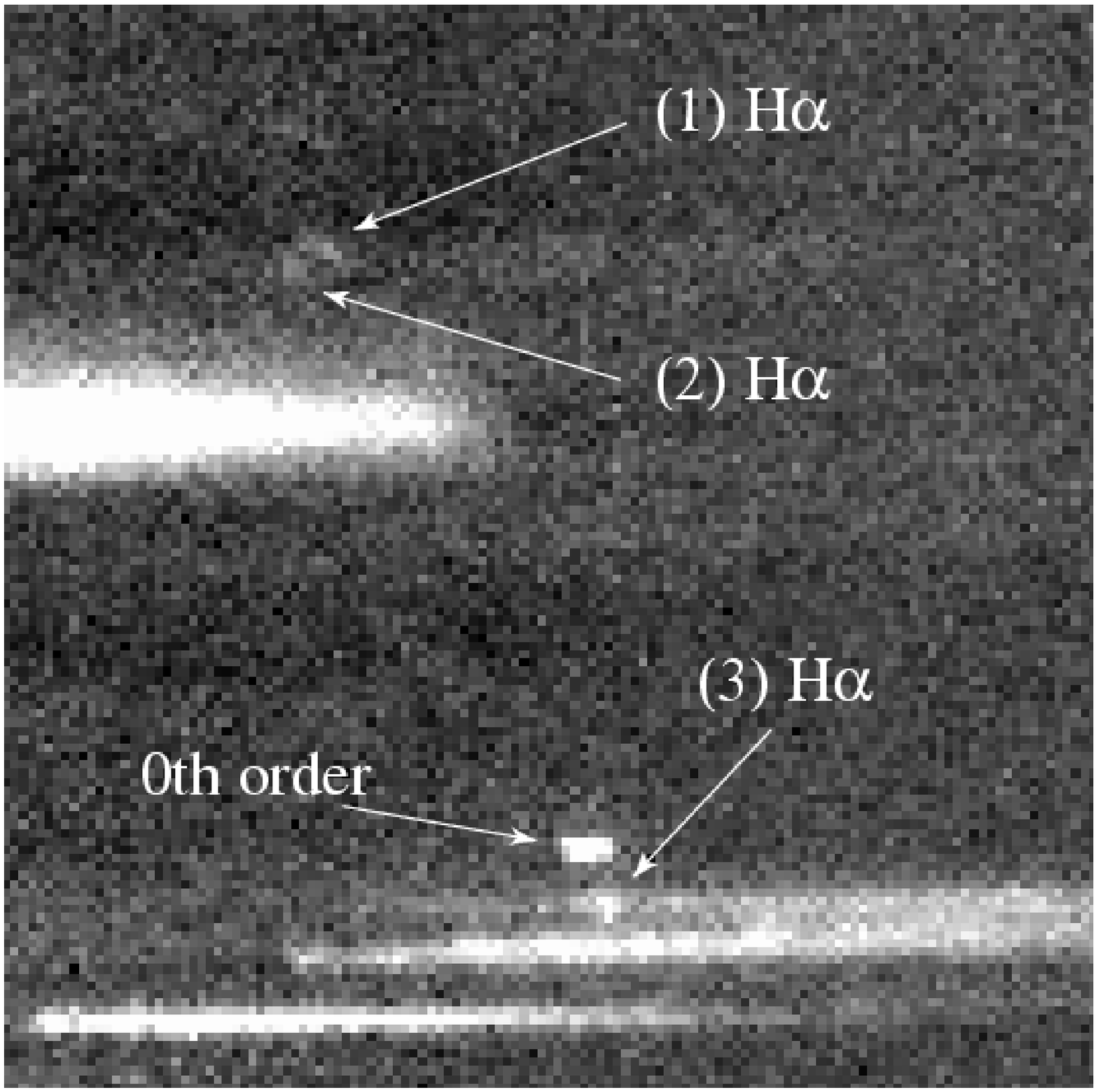}}}
\caption{Portions of NICMOS direct and grism images showing the galaxies
(1) NIC~J141725.66+522652.3, (2) NIC~J141725.71+522652.9 and
(3) NIC~J141727.34+522647.0. NIC~J141727.34+522647.0 is also visible in
a WFPC2 (F814W filter) image presented in \citet[ their Figure~7]{Gla:99},
close to the galaxy CFRS~14.1496. CFRS~14.1496 is the brighter object
immediately to the right of NIC~J141727.34+522647.0 in this image. The grism
image (right) indicates the location of the H$\alpha$ emission lines in the
spectra. The isolated bright feature in the grism image is a zero-order image
from an object lying outside the visible area of the frame.
 \label{imgs1}}
\end{figure*}

NIC~J141725.66+522652.3: A $3.5\sigma$ feature at $1.77\,\mu$m may
be H$\alpha$ at $z=1.70$. The NICMOS direct image of this object
(see Figure~\ref{imgs1}) appears to show two components. It is possible
that this is a close pair of galaxies, or a single disturbed system.
In addition, this galaxy is only separated by $1''$
from NIC~J141725.71+522652.9 and may be interacting with that galaxy
as well.

NIC~J141725.66+522724.7: This galaxy, identified with CFRS~14.1537
\citep{Lil:95}, has a very low S/N feature at $1.36\,\mu$m which may
be H$\alpha$ at $z=1.07$. No ground-based spectroscopic redshift
is available.

NIC~J141725.71+522652.9: A $6.5\sigma$ feature at $1.78\,\mu$m may
be H$\alpha$ at $z=1.71$. This galaxy may be interacting
with NIC~J141725.66+522652.3. It is possible that the observed H$\alpha$
emission may be related to any interaction, and estimated star formation
rates in both galaxies are high (assuming no AGN component to the source
of ionising radiation). See Figure~\ref{imgs1}.

NIC~J141726.75+522449.0: This galaxy is identified with object 104$-$4024
from \citet{Koo:96}. A very low S/N feature at $1.20\,\mu$m may be H$\alpha$
at $z=0.83$. The ground-based spectroscopic redshift of 0.8116 \citep{Koo:96}
is given a ``probable" confidence, and the discrepancy, which is of the
order of the grism spectrum resolution, may be explained by very low S/N of
the emission feature in the grism spectrum. The morphology
of this galaxy is given as ``O" by \citet{Koo:96}, classifying
it as ``round or elliptical, centrally concentrated, smooth, and
largely symmetrical," consistent with the morphology seen in the
NICMOS direct image.

NIC~J141727.28+522637.4: This galaxy is also catalogued as CFRS~14.1501
\citep{Bri:98}. A $2.5\sigma$ feature at $1.31\,\mu$m may be
H$\alpha$ at $z=1.00$. This is consistent with the CFRS confidence class 2
redshift of $z=0.989$
\citep[for the CFRS redshift confidence definitions, see][]{LeF:95}.
\citet{Bri:98} class the morphology for this galaxy as Irregular.

NIC~J141727.34+522647.0: A redshift of 1.54 for this galaxy, calculated
assuming the strong emission feature at $1.67\,\mu$m is H$\alpha$,
is confirmed by a ground-based optical spectrum (the DEEP consortium,
private communication). This galaxy is visible in a WFPC2 image of
CFRS~14.1496 presented by
\citet[ their Figure~7, the object directly beneath CFRS~14.1496]{Gla:99}.
The WFPC2 image shows that this object appears to consist of two components.
Given the proximity of the components and the presence of H$\alpha$ in
emission, it is probable that this is a pair of interacting galaxies, with
the observed H$\alpha$ emission possibly a product of interaction-induced
star formation \citep[ for example]{Ken:98}. See Figure~\ref{imgs1}.

NIC~J141727.42+522644.5: This galaxy is also catalogued as CFRS~14.1496
\citep{Lil:95,Bri:98}. The grism spectrum for this object at both observed
orientations is heavily blended with the spectra
of nearby objects, but a low S/N feature is visible at $1.25\,\mu$m, which,
if H$\alpha$, gives $z=0.91$ and is consistent with the CFRS confidence
class 3 value of $z=0.8990$. This object was also observed by
\citet{Gla:99}, who did not detect any H$\alpha$ emission and estimated an
upper limit of $L_{H\alpha}\sim7\times10^{34}\,$W, which is
consistent with our measurement of $L_{H\alpha}\approx3.8\times10^{34}\,$W.

NIC~J141740.65+522648.8: A $4\sigma$ feature at $1.26\,\mu$m may be
H$\alpha$ at $z=0.92$. This galaxy is about $2\farcs3$ from
15V20, a radio source detected in a deep 5\,GHz survey \citep{Fom:91}.
The positional offset is just larger than the the quoted 5\,GHz
$1\sigma$ positional errors, and it is possible the objects are
associated. The 5\,GHz flux density of 15V20 is $22.7\,\mu$Jy.

NIC~J141741.93+522645.2: Also catalogued as CFRS~14.1146
\citep{Lil:95,Bri:98}, this galaxy has a $4\sigma$ feature at $1.15\,\mu$m
which may be H$\alpha$ at
$z=0.75$. This is consistent with the ground-based redshift of $z=0.744$
(CFRS confidence class 3). \citet{Bri:98} give the morphology of the galaxy
as either a merger (visual classification) or an Sbc (automated classifier).

\begin{figure*}
\centerline{\hfill
\rotatebox{0}{\includegraphics[width=7cm]{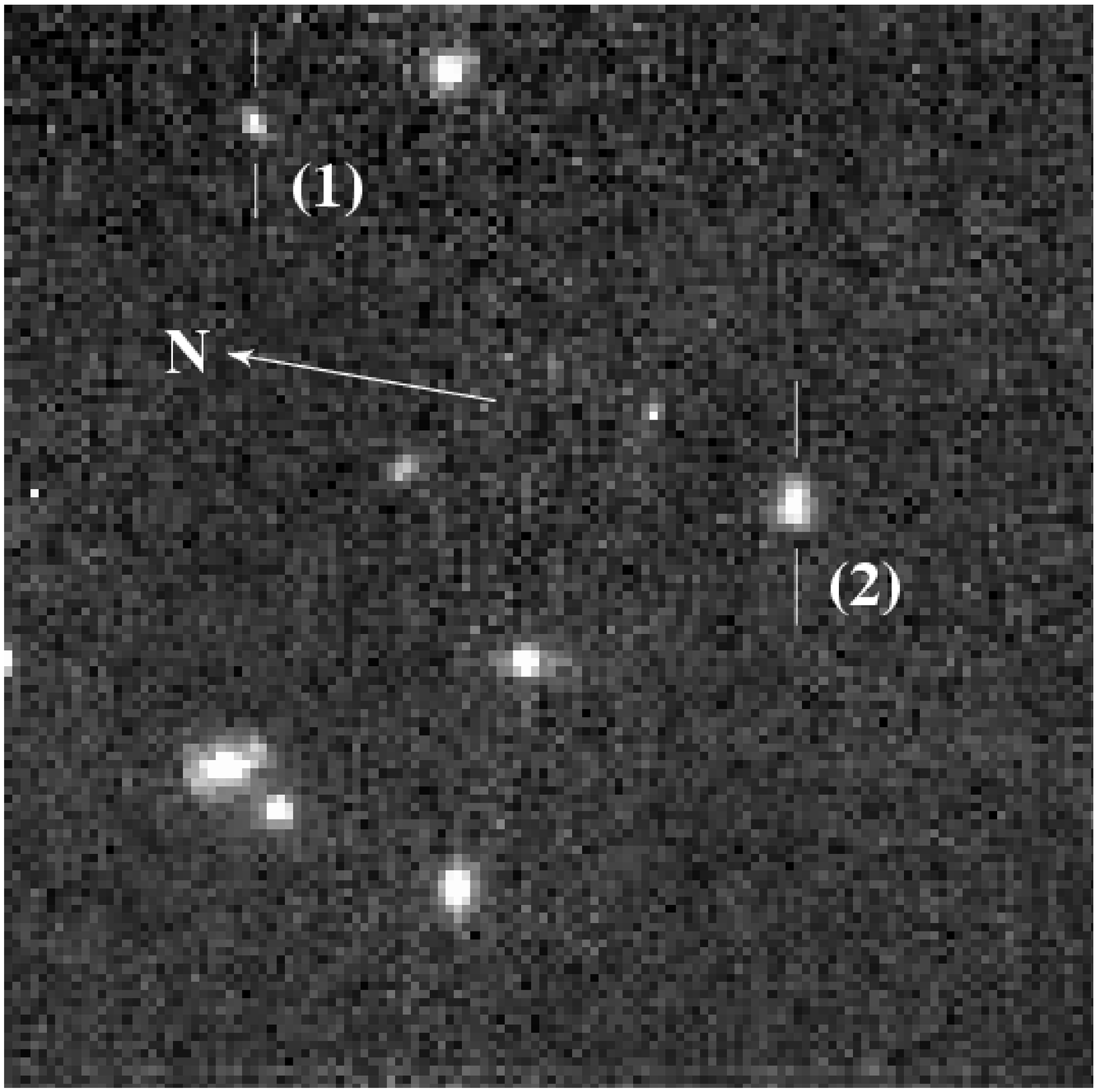}}
\hfill
\rotatebox{0}{\includegraphics[width=7cm]{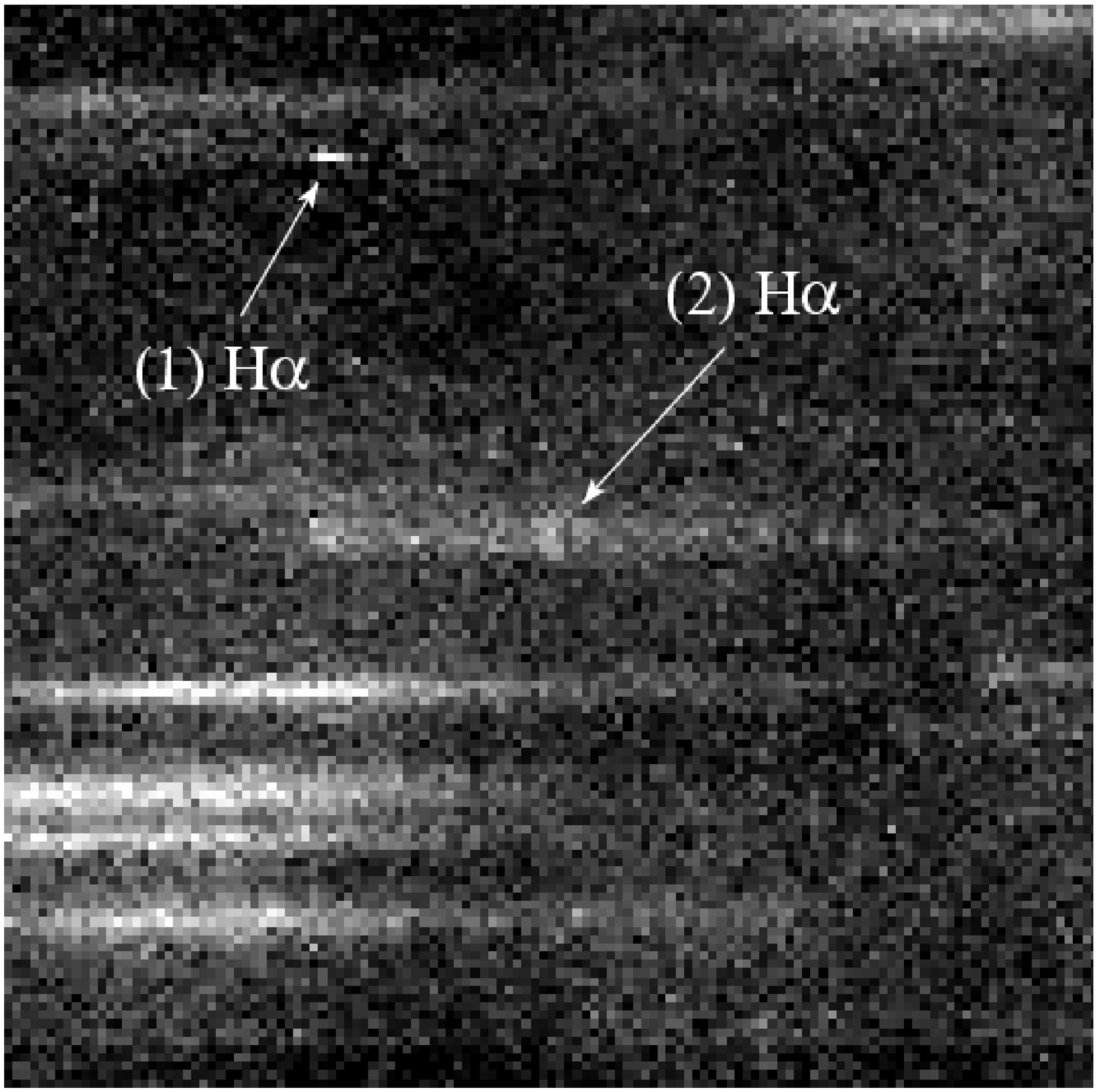}}}
\caption{Portions of NICMOS direct and grism images showing the galaxies
(1) NIC~J141750.32+523054.3 and (2) NIC~J141751.06+523040.0.
The H$\alpha$ emission features are again marked on the grism image.
 \label{imgs2}}
\end{figure*}

\begin{figure*}
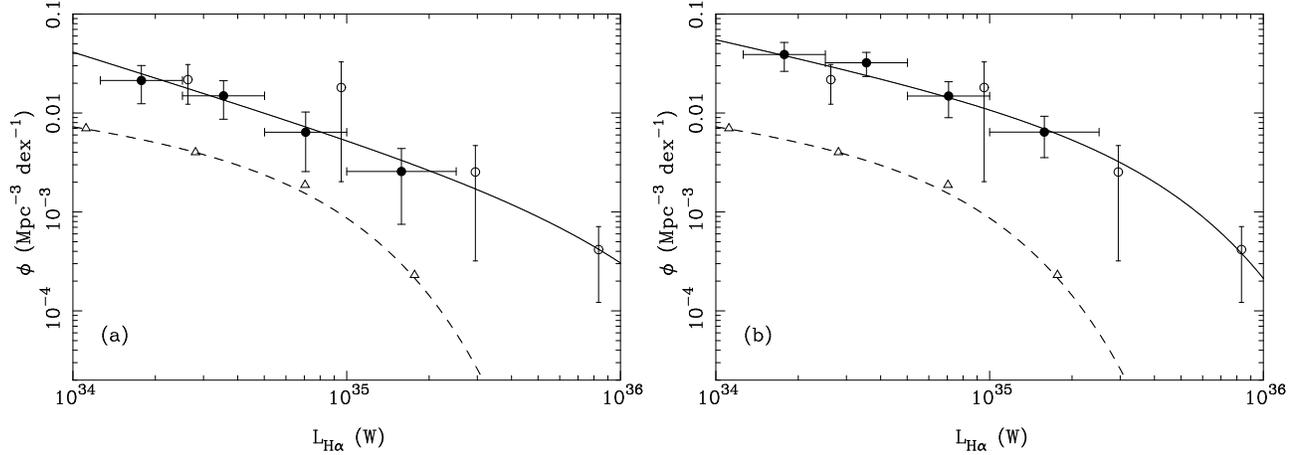

\centerline{\hfill
\rotatebox{-90}{\includegraphics[width=6cm]{hlf1.ps}}
\hfill
\rotatebox{-90}{\includegraphics[width=6cm]{hlf2.ps}}}
\caption{Derived H$\alpha$ luminosity functions (solid circles) along with
those of \protect\citet{Yan:99} (open circles). (a) LF derived only
from sources with spectroscopically confirmed redshifts or high S/N. (b) LF
derived from all sources with possible H$\alpha$ emission. The parameters of
the Schechter function fits are shown in Table~\protect\ref{schechpar}.
The triangles and dashed line in both diagrams show the local H$\alpha$
LF of \protect\citet{Gal:95} for comparison.
 \label{hlf}}
\end{figure*}

\begin{figure*}
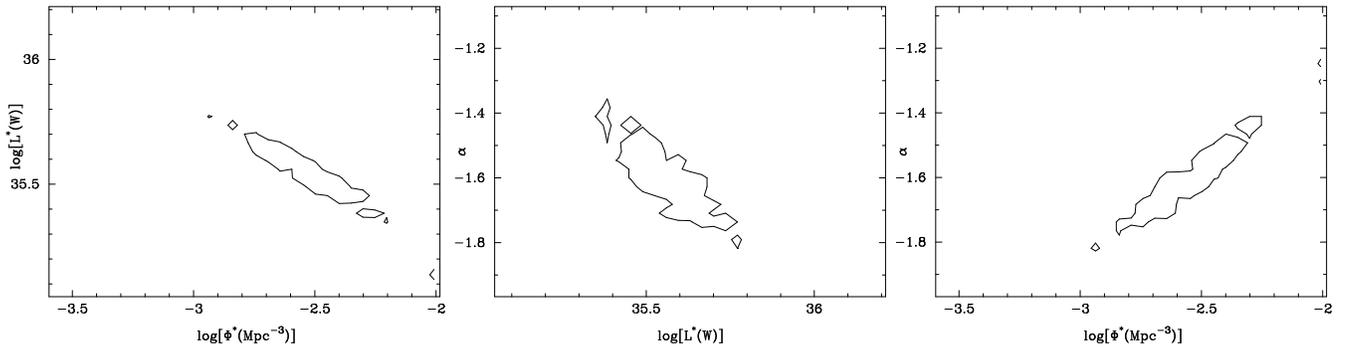

\centerline{\hfill
\rotatebox{-90}{\includegraphics[width=4.5cm]{pvsl.ps}}
\hfill
\rotatebox{-90}{\includegraphics[width=4.5cm]{lvsa.ps}}
\hfill
\rotatebox{-90}{\includegraphics[width=4.5cm]{pvsa.ps}}}
\caption{$1\,\sigma$ uncertainties displayed as ``error-areas"
for the Schechter function parameters, derived through the Monte-Carlo
method described in the text.
 \label{onesig}}
\end{figure*}

\begin{figure*}
\centerline{\rotatebox{-90}{\includegraphics[width=8cm]{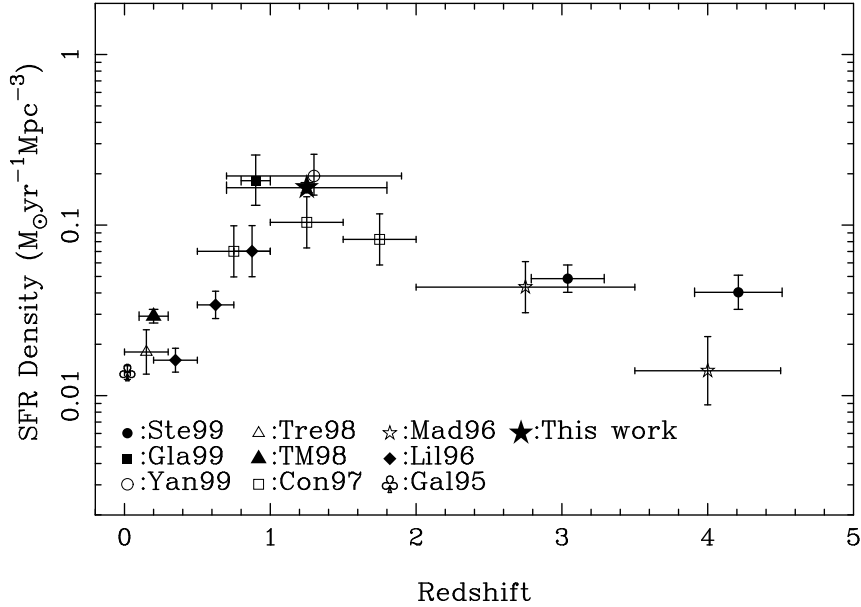}}}
\caption{SFR density as a function of redshift. This diagram is
a compilation of SFR densities derived from emission line and UV continuum
measurements taken from the literature. The UV based points
are shown here with no extinction corrections. Although omitted from this
diagram for reasons of clarity, the [O{\sc ii}] based estimates of
\citet{Hog:98} are also consistent with the H$\alpha$ estimates.
References in diagram are as follows, along with the origin of the SFR
density estimate:
Ste99: \citet{Ste:99} (UV); Gla99: \citet{Gla:99} (H$\alpha$);
Yan99: \citet{Yan:99} (H$\alpha$); Tre98: \citet{Tre:98} (UV);
TM98: \citet{TM:98} (H$\alpha$);
Con97: \citet{Con:97} (UV); Mad96: \citet{Mad:96} (UV);
Lil96: \citet{Lil:96} (UV); Gal95: \citet{Gal:95} (H$\alpha$).
 \label{sfd1}}
\end{figure*}

\begin{figure*}
\centerline{\rotatebox{-90}{\includegraphics[width=8cm]{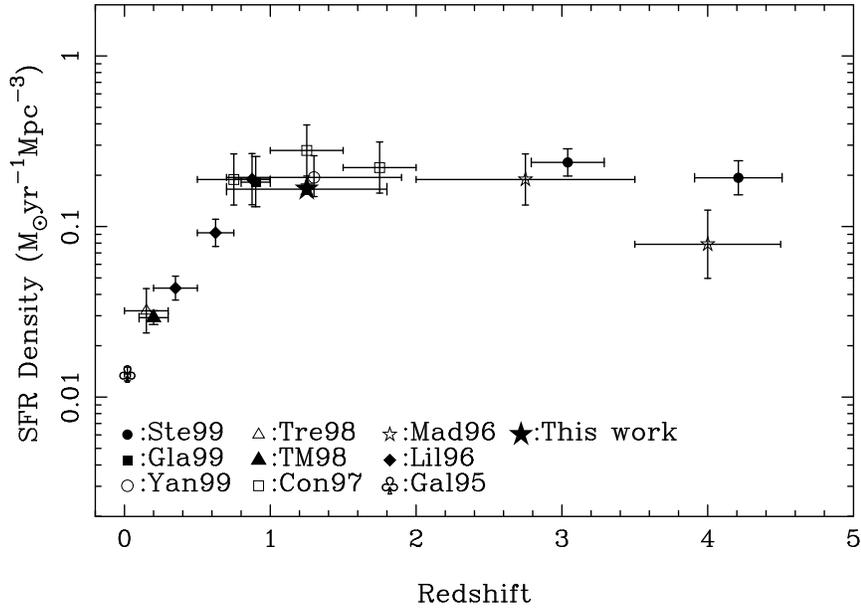}}}
\caption{SFR density as a function of redshift. This version
shows UV derived points with extinction corrections
as given by \protect\citet{Ste:99}. Symbols and references as in
Figure~\protect\ref{sfd1}.
 \label{sfd2}}
\end{figure*}

NIC~J141750.32+523054.3: This galaxy has a strong ($6\sigma$) feature
at $1.39\,\mu$m which may be H$\alpha$ at $z=1.12$. It is shown in
Figure~\ref{imgs2}.

NIC~J141751.06+523040.0: A $5.5\sigma$ feature at $1.67\,\mu$m may be
H$\alpha$ at $z=1.55$. It is shown in Figure~\ref{imgs2}.

NIC~J141751.99+523052.3: Also catalogued as CFRS~14.0846, this galaxy
shows a weak ($2\sigma$) feature at $1.31\,\mu$m which may be H$\alpha$ at
$z=1.00$. A ground-based optical spectrum \citep{Lil:95,Bri:98}
of this object gives a CFRS confidence class 2 redshift of
0.989, which would put H$\alpha$ at $1.305\,\mu$m, consistent within the grism
spectrum resolution. \citet{Bri:98} estimate the morphology for this galaxy as
Irregular.

NIC~J141752.29+523008.3: This galaxy is identified with object 063$-$2542
from \citet{Koo:96}. A $2\sigma$ feature at $1.38\,\mu$m may be H$\alpha$ at
$z=1.10$, consistent within the grism resolution of
the ground-based spectroscopic redshift of 1.0927. The morphology
of this galaxy is given as ``*" by \citet{Koo:96}, classifying
it as ``a galaxy likely to have on-going star formation with late-type,
irregular, asymmetrical, multicomponent, or peculiar forms."

NIC~J141753.47+522922.2: This galaxy has also been catalogued as CFRS~14.0807
\citep{Lil:95,Bri:98}. Its spectrum shows a broad but weak
($2\sigma$) feature at $1.31\,\mu$m which may be H$\alpha$ at $z=1.00$.
This is supported by a ground-based optical spectrum \citep{Bri:98} of
this object giving a CFRS confidence class 2 redshift of 0.985, which would put
H$\alpha$ at $1.30\,\mu$m. The discrepancy, still of the order of the
grism spectrum resolution, may be explained by combination of the low
S/N of the emission feature in the grism spectrum and its
broadness. \citet{Bri:98} estimate the morphology for this galaxy
as either Sab (visual) or Sbc (automated).

NIC~J141754.00+522923.9: A weak ($2\sigma$) feature at $1.31\,\mu$m may
be H$\alpha$ at $z=0.99$.
 This is supported by a ground-based optical
redshift for this object (the DEEP consortium, private communication).
 A disk is visible in the NICMOS direct image of
this galaxy, and it may be an early-type spiral.
Additionally, it is about $5''$ from NIC~J141753.47+522922.2
at a similar redshift. This corresponds to a separation of $\sim30\,$kpc.
They may be associated, possibly interacting.

NIC~J141754.29+523023.0: A weak ($2\sigma$) feature at $1.67\,\mu$m may
be H$\alpha$ at $z=1.43$. If so, an additional feature of similar
S/N at $1.19\,\mu$m would be [O{\sc iii}]$\lambda496\,$nm
emission.

NIC~J141754.35+523036.5: This galaxy has a photometric redshift
of $1.2\pm0.2$ from ground based photometry, consistent with the
location of the possible H$\alpha$ emission.

NIC~J141754.57+522930.6: A broad, weak emission feature ($2\sigma$)
at $1.31\,\mu$m may be a blend of H$\alpha$, [N{\sc ii}] and
the [S{\sc ii}]$\lambda\lambda672,673\,$nm doublet. Deblending the feature
(using {\sc splot}) gives a fit to the H$\alpha$-[N{\sc ii}] component
at $1.29\,\mu$m, implying $z=0.970$.
Within the limits of our resolution, this is consistent with a ground-based
optical redshift for this object (the DEEP consortium, private communication).
 
\section{Discussion}
\label{disc}

The conclusions of earlier work showing a strong increase of
global star formation rate out to redshifts of $\sim 1$ are clearly
seen in Figures~\ref{sfd1} and \ref{sfd2}. Our measured value
of 0.166\,M$_{\odot}$yr$^{-1}$Mpc$^{-3}$ for the star formation density
between $0.7<z<1.8$ reinforces this conclusion.
It also supports the conclusion that extinction corrections for UV derived
estimates of SFR density are significant, and need to be
of the magnitude applied by \citet{Ste:99} to ensure they are
consistent with estimates based on emission line measurements.

Our measurement of SFR density between $0.7<z<1.8$ is greater than that
estimated from UV based data by about a factor of $\approx 2-3$. This is
similar to the factor observed in the local universe and implies that,
globally speaking, there has been very little evolution in the relative
extinctions between UV and H$\alpha$ out to $z\approx1.3$. This in turn
implies that extinction laws constructed from observations of local systems
should be valid out to redshifts of at least $z=1$. Obviously individual
systems may show quite a wide variety in the amount or type of extinction,
however, and in general such global trends should be applied with caution
to individual cases.

Figure~\ref{sfd1} also emphasises that emission line measures
(O[{\sc ii}] and H$\alpha$) of SFR density are self-consistent, as are
the UV-based estimates, but that they are inconsistent with each other.
This discrepancy is most likely to be explained by dust extinction.
\citet{Ste:99} has shown that the extinction law of \citet{Cal:97a,Cal:97b} for
starburst regions appears to be quite successful in accounting for this
discrepancy.

\section{Conclusion}
\label{conc}

We have observed 6 fields lying within the Groth-Westphal strip with
the NICMOS G141 grism, covering a total area of $\sim 4.4$ square arcminutes.
A catalogue of 163 objects has been compiled, 37 of which show emission
features likely to be H$\alpha$ at redshifts between $0.7<z<1.8$. The
H$\alpha$ LF in this range has been extended to luminosities a factor of two
fainter than measured by \citet{Yan:99}. We find a value for the density of
star formation between $0.7<z<1.8$ of 0.166\,M$_{\odot}$yr$^{-1}$Mpc$^{-3}$,
consistent with earlier H$\alpha$ measurements. This supports the
conclusions of earlier work that the global star formation rate
at $z\approx1.3$ is an order of magnitude greater than locally,
and that extinction corrections for UV derived estimates of SFR density
are significant. Additionally, our measurement of SFR density at
$z\approx1.3$ is greater than the UV estimate by a similar factor to that
found locally. This implies little, if any, cosmic evolution in the relative
extinctions between UV and H$\alpha$ out to this redshift.

\acknowledgements

The authors would like to thank the DEEP consortium for providing
data ahead of publication for comparison with our results.
We also thank the referee for helpful comments.
This research is based on observations with the NASA/ESA Hubble Space
Telescope, obtained at the Space Telescope Science Institute, which is
operated by the Association of Universities for Research in Astronomy, Inc.
under NASA contract No. NAS5-26555. Support for this work was provided by NASA
through grant number GO-07871.02-96A from the Space Telescope Science
Institute, which is operated by AURA, Inc., under NASA contract NAS5-26555.
AJC and AMH also acknowledge partial support from NASA LTSA grant
NRA-98-03-LTSA-039.
This research has made use of the NASA/IPAC Extragalactic Database (NED)
which is operated by the Jet Propulsion Laboratory, California Institute of
Technology, under contract with the National Aeronautics and Space
Administration.

\end{document}